\documentstyle[12pt]{article}
\begin{document}
\begin{center}{\Large {\bf Inflection point as a manifestation of tricritical point on
the dynamic phase boundary in Ising meanfield dynamics}}\end{center}

\vskip 0.5cm

\begin{center}{\it Muktish Acharyya}\\

{Department of Physics, Presidency College}\\

{86/1 College Street, Calcutta-700073, India}\\

{\it {E-mail: muktish@vsnl.net}}\\

{and}\\

{\it Ajanta Bhowal Acharyya}\\

{Department of Physics, Hooghly Mohsin College}\\

{PO-Chinsurah, Dist-Hooghly, India}\end{center}

\vskip 1 cm

\noindent Abstract: We studied the dynamical phase transition in kinetic Ising
ferromagnets driven by oscillating magnetic field in meanfield approximation. The meanfield
differential equation was solved by sixth order Runge-Kutta-Felberg method. We
calculated the transition temperature as a function of amplitude and frequency of
oscillating field. This was plotted against field amplitude taking frequency as a
parameter. As frequency increases the phase boundary is observed to become inflated. 
The phase boundary shows an inflection point which separates
the nature of the transition.
On the dynamic phase boundary a tricritical point (TCP) was found, which separates the
nature (continuous/discontinuous) of the dynamic transition across the phase boundary.
The inflection point is identified as the TCP and hence
a simpler method of determining the position of TCP was found. TCP was observed
to shift towards high field for higher frequency.  
As frequency decreases the dynamic phase boundary is observe to shrink.
In the zero frequency limit this boundary shows a tendency to 
merge to the temperature variation of the coercive field. 

\vskip 0.5cm


\newpage

\noindent {\bf I. Introduction:}

The ferromagnetic system, in the presence of a time varying external magnetic field, remaining far from
statistical equilibrium, became an interesting object of research over the last two decades \cite{rev}.
One interesting nonequilibrium response is the dynamic phase transition. This dynamic phase transition
is widely studied in model ferromagnetic system in the presence of oscillating magnetic field \cite{rev}.
Tome and Oliveira \cite{tom} first observed a prototype of nonequlibrium dynamic transition in the 
numerical solution of meanfield equation of motion for the classical Ising ferromagnet in the presecce of
a magnetic field varying sinusoidally in time. The time averaged (over the complete cycle of the oscillating
magnetic field) magnetisation plays the role of the dynamic order parameter. They \cite{tom} found that this
dynamic ordering depends on the amplitude of the oscillating magnetic field and the temperature of the system.
Systems get dynamically ordered for small values of the temperature and the amplitude of the field. They \cite{tom}
have drawn a phase boundary (separating the ordered and disordered phase) in the temperature field amplitude
plane. More interestingly, they have also reported \cite{tom} a tricritical point on the phase boundary, which
separates the nature (continuous/discontinuous) of the dynamic transition across the phase boundary. This tricritical
point was found just by checking the nature of the transition at all points across the phase boundary. The point where
the nature of transition changes was marked as the tricritical point. No other significance of this tricritical 
point was reported. 
The frequency dependence of this phase boundary was not reported earlier for the dynamic transition in Ising
meanfield dynamics.

In this paper, we studied numerically the dynamic transition in Ising meanfield dynamics. Here, we confined our
attention to study the frequency dependence of the dynamic phase boundary. We studied the tricritical behaviour
and found a method of finding the position the tricritical point on the dynamic phase boundary. 
The frequency dependence of the position of the tricritical point was studied here. We also studied the static
(zero frequency) limit of dynamic phase boundary.

The paper is organised as follows: In the next section the model and the method of numerical solution is discussed.
Section III contains the numerical results and the paper end with summary of the work in section IV.

\newpage

\noindent {\bf II. Model and numerical solution:}

The time ($t$) variation of average magnetisation $m$ of Ising ferromagnet in the
presence of a time varying field, in meanfield approximation, is given as \cite{tom}

\begin{equation}
\tau {{dm} \over {dt}} = -m + {\rm tanh}({{m+h(t)} \over {T}}),
\end{equation}

\noindent where, $h(t)$ is the externally applied sinusoidally oscillating
magnetic field ($h(t) = h_0 {\rm sin}(\omega t)$) and $T$ is the temperature measured
in units of the Boltzmann constant ($K_B$). This equation describes the nonequilibrium
behaviour of instantaneous value of magnetisation $m(t)$ of Ising ferromagnet in 
meanfield approximation.

We have solved this
equation by sixth order Runge-Kutta-Felberg (RKF) \cite{numeric} method to get the instantaneous value
of magnetisation $m(t)$ at any finite temperature $T$, $h_0$ and $\omega (=2\pi f)$.
This method of solving ordinary differential equation ${{dm} \over {dt}} = F(t,m(t))$, is
described briefly as:
\begin{flushleft}{$m(t+dt) = m(t) + \left({{16k_1} \over {135}}+{{6656k_3} \over {12825}}
+{{28561k_4} \over {56430}}-{{9k_5} \over {50}}
+{{2k_6} \over {55}}\right)$}\\
{\rm where}\\
{$k_1 = dt \cdot F(t,m(t))$}\\
{$k_2 = dt \cdot F(t+{{dt} \over 4}, m+{{k_1} \over 4})$}\\
{$k_3 = dt \cdot F(t+{{3dt} \over 8}, m+{{3k_1} \over {32}}+{{9k_2} \over 32})$}\\
{$k_4 = dt \cdot F(t+{{12dt} \over {13}}, m+{{1932k_1} \over {2197}}
-{{7200k_2} \over {2197}}+{{7296k_3} \over 2197})$}
{$k_5 = dt \cdot F(t+dt,m+{{439k_1} \over 216}-8k_2+{{3680k_3} \over 513}
-{{845k_4} \over 4104})$}\\
{$k_6 = dt \cdot F(t+{{dt} \over 2}, m-{{8k_1} \over 27}+2k_2-{{3544k_3} \over 2565}
+{{1859k_4} \over 4104}-{{11k_5} \over 40})$.................................(2)}
\end{flushleft}

\noindent The time interval $dt$ was measured in units of 
$\tau$ (the time taken to flip a single
spin). Actually, we have used $dt = 0.01$ (setting $\tau$=1.0). 
The local error involved in the sixth order RKF method
is of the order of $(dt)^6 (=10^{-12})$. We started with initial condition $m(t=0) = 1.0$. 

\vskip 1cm

\noindent {\bf III. Results:} 

The dynamic order parameter $Q (= {{2\pi} \over {\omega}}
\oint m(t) dt)$ is time average magnetisation over a full cycle of the oscillating
magnetic field. This was calculated after discarding the 
values of $Q$ for few initial (transient \cite{mapre1}) cycles of the
oscillating field. Finally, the dynamic order parameter $Q$ is calculated as a function
of $T$, $h_0$ and $f$. Now, depending on the values of these parameters the
system gets dynamically ordered ($Q \neq 0$) or disordered ($Q = 0$). This shows a dynamical
phase transition which is a nonequilibrium phase transition. We have studied
the transition and determined the transition temperature $T_d(h_0,f)$ in a very simple way. 
For a fixed set of values of $h_0$ and $f$
the temperature $T$ is varied (in step $\Delta T = 10^{-3}$) and $Q$ is measured as a
function of $T$. Then we calculated the derivative ${{dQ} \over {dT}}$ numerically
(using three point central difference formula; where the error  $O(dT^2)$ \cite{numeric}). 
The temperature, at which the derivative ${{dQ} \over {dT}}$ is sharply minimum,
is considered here as the transition temperature $T_d$. 
In this way, we obtained the dynamic transition temperature
$T_d(h_0,f)$ for all values of $h_0$. Here, we have changed $h_0$ (with interval $\Delta h_0
= 0.02$) and obtained the dynamic transition temperature $T_d(h_0,f)$.

Now, for a particular frequency $f$, 
the plot of $T_d(h_0,f)$ against $h_0$
gives the dynamic phase boundary. This dynamic phase boundary separates the regions of
$Q \neq 0$ and $Q = 0$. For fixed frequency, it was observed that the dynamic transition
occurs at higher temperature for lower values of applied field amplitude $h_0$ and vice
versa. Fig.1 shows such a variation. For $f=0.2$ and $h_0=0.5$, 
the temperature variations of $Q$ and 
${{dQ} \over {dT}}$ are plotted in Fig.1(A) and Fig.1(B) respectively. From the sharp
minimum of ${{dQ} \over {dT}}$ (in Fig.1(B)) the transition temperature $T_d(h_0,f)$
was found equal to 0.725. The same plots are shown in Fig.1(C) and (D) for $h_0=0.3$
(keeping frequency $f=0.2$ fixed). Here, the transition temperature was found to be
equal to 0.919. 
It is clear from the figure that the transition occurs at lower temperature for higher
value of the field amplitude.
In this way, the entire phase boundary 
(i.e., $T_d(h_0,f)$ as a function of $h_0$ for fixed $f=0.2$) was obtained.

The dynamic phase boundary was 
obtained for different frequency $f$. It was observed that for a fixed value of the
field amplitude $h_0$ that transition occurs at higher temperature for higher frequency.
This observation was shown in Fig.2, for fixed $h_0=0.4$. The dynamic transition occurs
at $T_d(h_0,f)=0.845$ for $f=0.2$ (see Fig.2(A) and (B)) 
and it becomes $T_d(h_0,f)=0.908$ for $f=1.0$ (see Fig.2(C) and (D)). From the figures it is 
clear that the transition occurs at higher temperature for higher frequency. We have reported
the results of dynamic phase boundary for frequencies $f$ = 0.01, 0.02, 0.05, 0.1, 0.2, 0.5
and 1.0.

In the
limit $f \to 0$ we approach the equilibrium behaviour. In equilibrium, ${{dm} \over {dt}} 
=0$. Equation (1) takes the form $m - {\rm tanh}{{(m+h)} \over T} = 0$. This equation was solved
by Newton-Raphson iterative method of finding the 
root to get the value of equilibrium magnetisation $m(h,T)$.
At any fixed temperature $T$, by changing
the value of the external magnetic field $h$, we calculated the coercive field $h_c$ (i.e,
the field for which the magnetisation just changes its sign). The value of the coercive
field was found to depend on the temperature $T$, i.e.,  the  coercive field is a function
of the temperature.

The dynamic transition temperature $T_d$ is also
plotted against the amplitude of the externally applied sinusoidal magnetic field, in the
same figure. For a fixed value of frequency, the transition temperature $T_d$ decreases as
the value of the field amplitude increases. This gives the dynamic phase boundary, below
which we observed the dynamically ordered ($Q \neq 0$) phase and above which the phase is
dynamically disordered ($Q = 0$). Different dynamic phase boundary was obtained for different
values of frequency and plotted in Fig.3. It is observed that for same value of the field
amplitude, the transition temperature $T_d$ increases as the frequency increases. So, the
phase boundary gets inflated as the frequency increases. 

As the frequency decreases ($f \to 0$) the
phase boundary shrinks and ultimately it approches the curve of temperature variation
of the coercive field (continuous line in Fig.3). 
In this context, one may think that the temperature variation of
coercive field acts as the static limit of dynamic phase boundary. We studied the nature
of the dynamic phase boundary. This dynamic phase boundary changes its 
curvature from one side to other, as
one changes the field from lower to higher value. This means, the boundary has an inflection
point (where the curvature changes from one side of the curve to its other side) 
which was detected
by calculating the derivative ${{dT_d} \over {dh_0}}$. The derivative ${{dT_d} \over {dh_0}}$
plotted against $h_0$ (for fixed $f=1.0$) shows (in Fig.4) a very sharp minimum (at $h_m$) 
which indicates the inflection point of $T_d-h_0$ curve.
This minimum occurs at $h_0=0.72$.
This minimum or the inflection point has a great significance. 
If the value of the field amplitude 
is less than $h_m$, the transition is a continuous one. 
A typical transition, for $h_0 = 0.70 (< h_m)$, was shown in the left inset of Fig.4. 
On the other hand, if the value
of the field amplitude exceeds $h_m$, the transition becomes a discontinuous
one. The right inset of Fig.4,
shows such a typical transition for $h_0 = 0.74 (> h_m)$. The observation shows that
the inflection point on the dynamic phase boundary acts as tricritical point.
To search the location of the tricritical point on the dynamic phase boundary, the nature of
the dynamic transition has to be studied at several points. The TCP is the point, where
the nature (continuous/discontinuous) of the transition changes from one side to other. 
But in the method of finding the inflection point on the dynamic phase boundary,
one can get the exact location of tricritical point on the phase boundary, quite easily
at least in this case. 
Tome and Oliveira \cite{tom} studied
this dynamic transition in kinetic Ising model in meanfield approximation and observed
the existence of a tricritical point on the phase boundary. But they did not report any
method to find the TCP directly from the phase boundary. Here, we found a method of 
getting the TCP directly from the phase boundary.

We have also studied the change in position of the TCP on the phase boundary as one changes
the frequency. Following the same method, by computing the derivative ${{dT_d} \over {dh_0}}$
and plotting it against $h_0$, for different frequency, we obtained the position of the
minima of the derivatives (for different frequencies) and hence the position of TCP's. This
was shown in Fig.5 for few frequencies. The position of TCP's for different frequencies are
shown (by big black dot) in Fig.3. It was observed that the position of TCP shifts towards
higher field amplitudes (consequently lower temperatures) for higher frequencies.

\vskip 0.5cm

\noindent {\bf IV. Summary:}

In this paper, we have reported our numerical results of the study of the dynamic
phase transition in kinetic Ising model driven by oscillating magnetic field, in the
meanfield approximation. The dynamic phase boundary was drawn in the temperature-field amplitude
plane for different frequencies of the applied oscillating magnetic field. The dynamic phase
boundary was observed to get inflated as the frequency increases. As the frequency decreases
it shrinks and in zero frequency limit it seems to merge to the temperature variation
of the coercive field (equilibrium case).

The important thing that we observed here is:  The tricritical point on the dynamic phase
boundary is the point of inflection of the phase boundary. This observation made the
task, of finding the position of TCP on the phase boundary, much simpler. We have also
observed that the position of TCP shifts towards higher field amplitude for higher frequency.

\noindent {\bf Acknowledgments:} The library facility provided by Saha Institute of Nuclear
Physics, Calcutta, is gratefully acknowledged.

\newpage
\topmargin=-1cm
\oddsidemargin=4cm
\setlength{\unitlength}{0.240900pt}
\ifx\plotpoint\undefined\newsavebox{\plotpoint}\fi
\sbox{\plotpoint}{\rule[-0.200pt]{0.400pt}{0.400pt}}%


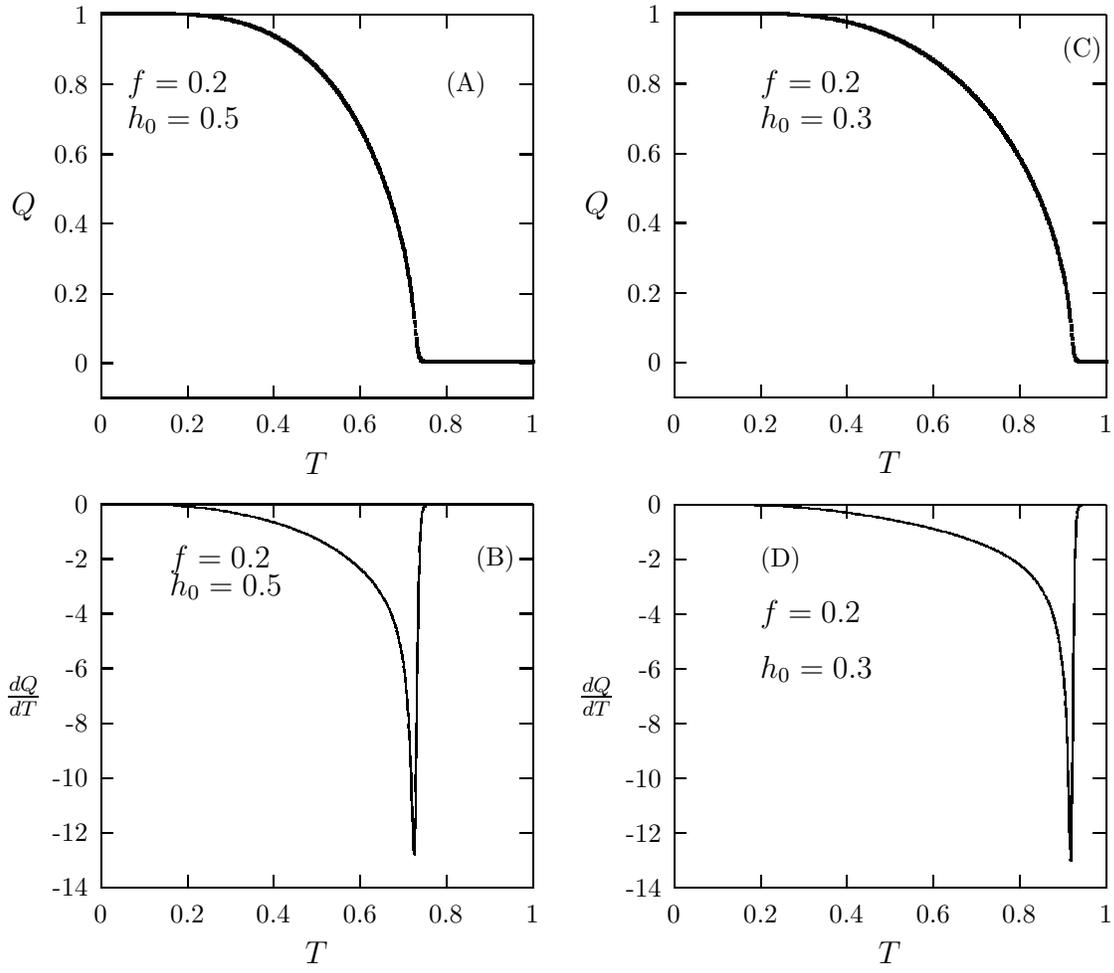
\noindent {\bf Fig.1}. The temperature variation of $Q$ and ${{dQ} \over {dT}}$ for different values of $h_0$ but for fixed $f=0.2$.
(A) and (B) for $h_0=0.5$ and (C) and (D) for $h_0=0.3$.

\newpage

\topmargin=-1cm
\oddsidemargin=4cm
\setlength{\unitlength}{0.240900pt}
\ifx\plotpoint\undefined\newsavebox{\plotpoint}\fi
\sbox{\plotpoint}{\rule[-0.200pt]{0.400pt}{0.400pt}}%
%

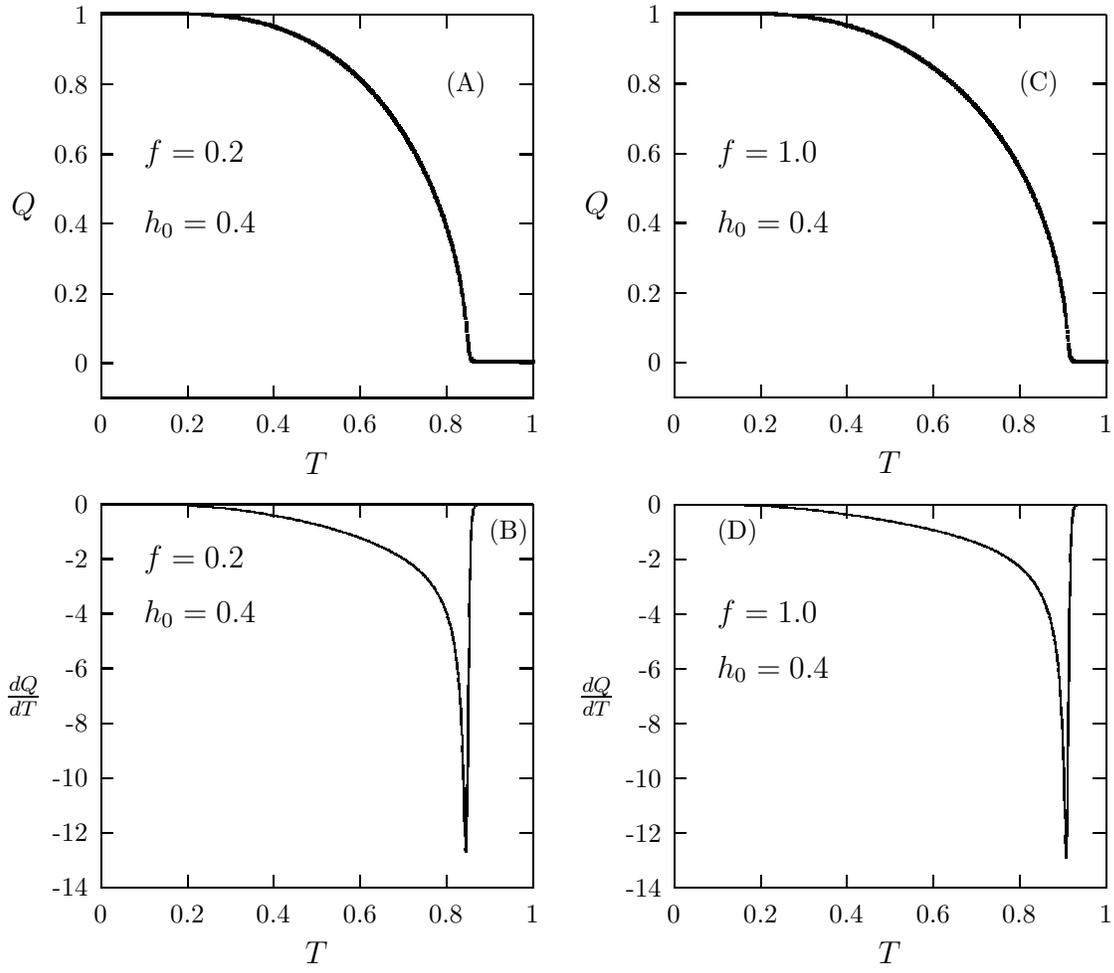
\noindent {\bf Fig.2}. The temperature variation of $Q$ and ${{dQ} \over {dT}}$ for different values of $f$ but for fixed $h_0=0.4$.
(A) and (B) for $f=0.2$ and (C) and (D) for $f=1.0$.
\newpage

\setlength{\unitlength}{0.240900pt}
\ifx\plotpoint\undefined\newsavebox{\plotpoint}\fi
\sbox{\plotpoint}{\rule[-0.200pt]{0.400pt}{0.400pt}}%
%

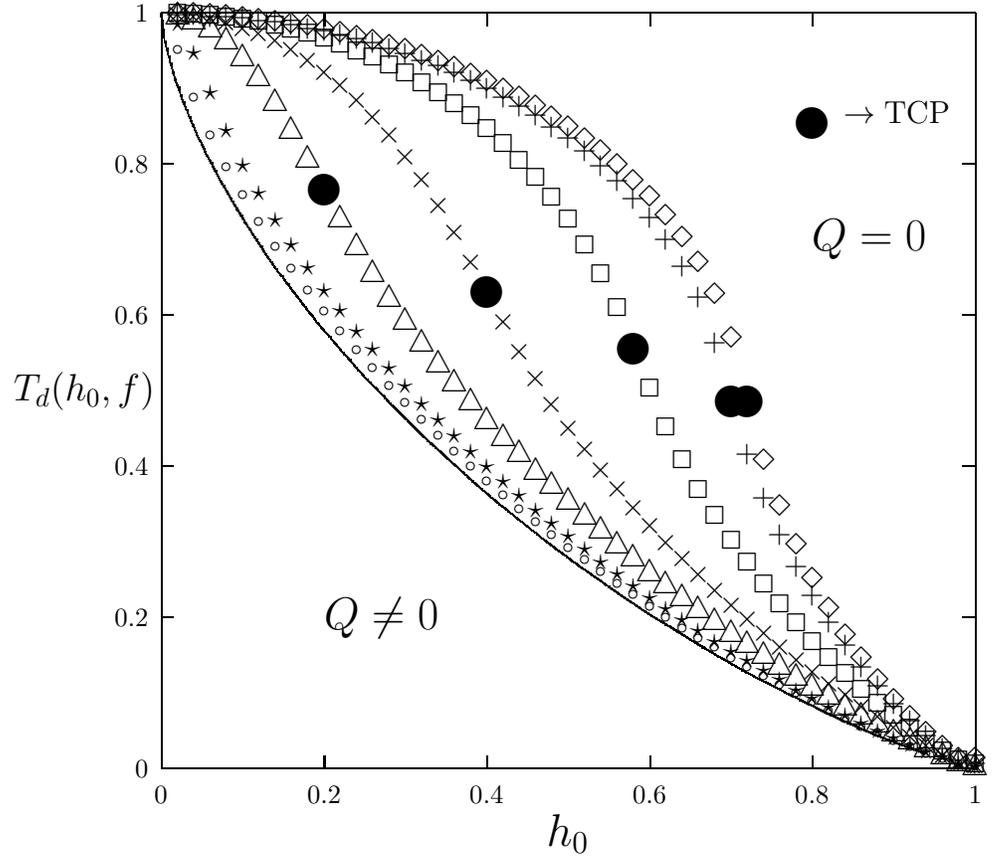
\noindent {\bf Fig.3}. The dynamic transition temperature $T_d(h_0,f)$ is plotted against the amplitude of oscillating
magnetic field $h_0$ taking frequency $f$ as parameter. Different symbols represent different frequencies. 
$f=1.0(\Diamond)$, $f=0.5(+)$, $f=0.2(\Box)$, $f=0.1({\times})$, $f=0.05({\triangle})$, $f=0.02({\star})$ and $f=0.01(o)$.
The bullets represent the TCP's on the phase boundary. 
The continuous line represent the variation of coercive field with temperature.

\newpage

\setlength{\unitlength}{0.240900pt}
\ifx\plotpoint\undefined\newsavebox{\plotpoint}\fi
\sbox{\plotpoint}{\rule[-0.200pt]{0.400pt}{0.400pt}}%


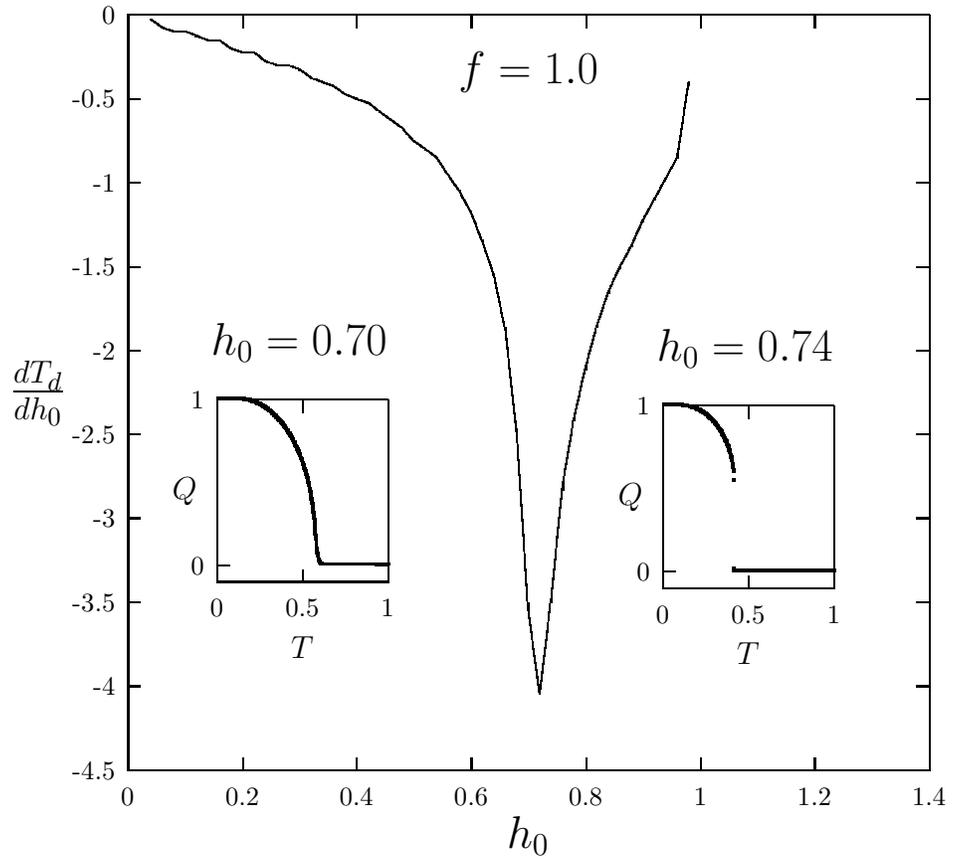
\noindent {\bf Fig.4}. The derivative ${{dT_d} \over {dh_0}}$ is plotted against the field amplitude $h_0$ for a
particular frequency $f=1.0$. Left inset shows the temperature variation of the dynamic order parameter for
$h_0=0.70$ and the right inset shows that for $h_0=0.74$. 

\newpage

\setlength{\unitlength}{0.240900pt}
\ifx\plotpoint\undefined\newsavebox{\plotpoint}\fi
\sbox{\plotpoint}{\rule[-0.200pt]{0.400pt}{0.400pt}}%
%

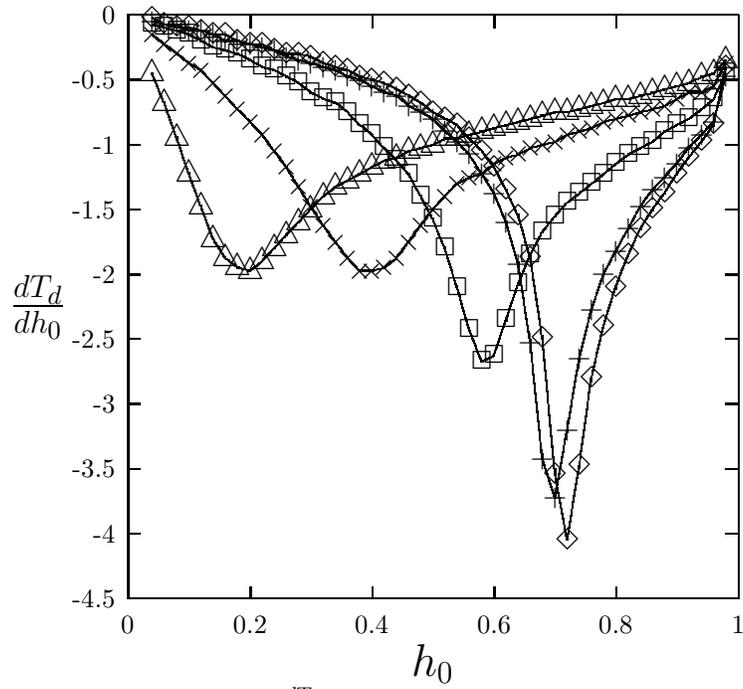
\noindent {\bf Fig.5}. The derivative ${{dT_d} \over {dh_0}}$ is plotted against the field amplitude $h_0$ for 
different frequencies. 
Different symbols represent different frequencies. 
$f=1.0(\Diamond)$, $f=0.5(+)$, $f=0.2(\Box)$, $f=0.1({\times})$ and $f=0.05({\triangle})$.
The continuous 
line in each case just connects the data points.
\end{document}